\documentclass[12pt]{article}
\usepackage{amsmath,amssymb,epsfig}
\begin{document}
\setlength{\parskip}{0.45cm}
\setlength{\baselineskip}{0.75cm}
%
%
%
\begin{titlepage}
\setlength{\parskip}{0.25cm}
\setlength{\baselineskip}{0.25cm}
\begin{flushright}
DO-TH 2004/13\\
\vspace{0.2cm}
December 2004
\end{flushright}
\vspace{1.0cm}
\begin{center}
\Large
{\bf Probing the Perturbative NLO Parton Evolution}
\\\Large{\bf in the Small--$x$ Region}
\vspace{1.5cm}

\large
M.\ Gl\"uck, C.\ Pisano, E.\ Reya\\
\vspace{1.0cm}

\normalsize
{\it Universit\"{a}t Dortmund, Institut f\"{u}r Physik,}\\
{\it D-44221 Dortmund, Germany} \\
\vspace{0.5cm}

\vspace{1.5cm}
\end{center}

\begin{abstract}
\noindent A dedicated test of the perturbative QCD NLO parton evolution
in the very small--$x$ region is performed.  We find a good
agreement with recent precision HERA--data for $F_2^p(x,Q^2)$,
as well as with the present determination of the curvature of $F_2^p$.
Characteristically, perturbative QCD evolutions result in a 
{\underline{positive}} curvature which increases as $x$ decreases.
Future precision measurements in the very small $x$--region,
$x<10^{-4}$, could provide a sensitive test of the range of 
validity of perturbative QCD.
\end{abstract}
\end{titlepage}


Parton distributions $f(x,Q^2)$, $f=q,\, \bar{q},\, g$, underlie a
$Q^2$--evolution dictated by perturbative QCD at 
$Q^2$ \raisebox{-0.1cm}{$\stackrel{>}{\sim}$} 1 GeV$^2$.
It was recently stated \cite{ref1} that the NLO perturbative
QCD $Q^2$--evolution disagrees with HERA data \cite{ref2,ref3}
on $F_2^p(x,Q^2)$ in the small--$x$ region, 
\mbox {$x$ \raisebox{-0.1cm}{$\stackrel{<}{\sim}$} $10^{-3}$.}  
In view
of the importance of this statement we perform here an independent
study of this issue.  In contrast to \cite{ref1} we shall undertake
this analysis in the standard framework where one sets up input
distributions at some low $Q_0^2$, here taken to be $Q_0^2 = 1.5$
GeV$^2$, corresponding to the lowest $Q^2$ considered in 
\cite{ref1}, and adapting these distributions to the data
considered.  In the present case the data considered will be
restricted to 
\begin{equation}
1.5\,\,{\rm GeV}^2\leq Q^2\leq 12\,\,{\rm GeV}^2,\quad\quad
3\times 10^{-5} \lesssim x \lesssim 3\times 10^{-3}
\end{equation}
as in \cite{ref1} and will be taken from the corresponding
measured $F_2^p(x,Q^2)$ of the H1 \mbox{collaboration} \cite{ref2}.
The choice of these data is motivated by their higher precision
as compared to corresponding data of the ZEUS collaboration
\cite{ref3}, in particular in the very small--$x$ region.

We shall choose two sets of input distributions based on the
GRV98 parton distributions \cite{ref4}.  In the first set we
shall adopt $u_v,\, d_v,\, s=\bar{s}$ and $\Delta\equiv\bar{d}-
\bar{u}$ from GRV98 and modify $\bar{u}+\bar{d}$ and the gluon
distribution in the small--$x$ region to obtain an optimal fit
to the H1 data \cite{ref2} in the aforementioned kinematical
region.  We shall refer to this fit as the `best fit'.
The second choice will be constrained to modify the GRV98 
$\bar{u}+\bar{d}$ and $g$ distributions in the small--$x$ region
as little as possible.  We shall refer to this fit as  
${\rm GRV}_{\rm mod}$.  It will turn out that both input 
distributions are compatible with the data to practically the
same extent, i.e.\ yielding comparable $\chi^2/dof$. In view
of these observations we do not agree with the conclusions of
ref.~\cite{ref1}, i.e.\ we do not confirm a disagreement between
the NLO $Q^2$--evolution of $f(x,Q^2)$ and the measured
\cite{ref2,ref3} $Q^2$--dependence of $F_2^p(x,Q^2)$. 

The remaining flavor--singlet input distributions at $Q_0^2=1.5$
GeV$^2$ to be adapted to the recent small--$x$ data are expressed
as
\begin{eqnarray}
xg(x,Q_0^2)& = & N_g x^{-a_g}\left( 1 + A_g\sqrt{x}+7.283x\right)
 (1-x)^{4.759}\\
x(\bar{u}+\bar{d})(x,Q_0^2) & = & N_sa^{-a_s}
  \left(1+A_s\sqrt{x}-4.046x\right) (1-x)^{4.225}
\end{eqnarray}
where the parameters relevant for the large $x$--region, $x>10^{-3}$,
which is of no relevance for the present small--$x$ studies, are
kept unchanged and are taken from, e.g.\ GRV98 \cite{ref4}. The
refitted relevant small--$x$ parameters turn out to be
\begin{eqnarray}
{\rm `best}\,{\rm fit'} &:& N_g=1.70\, ,\,\, \quad a_g=0.027\, ,
        \quad A_g=-1.034\nonumber\\
& & N_s=0.171\, ,\quad a_s=0.177\, ,\quad A_s=2.613
\end{eqnarray}
\begin{eqnarray}
{\rm GRV}_{\rm mod} &:& N_g=1.443\, ,\quad a_g=0.125\, ,\quad
     A_g=-2.656\nonumber\\
& & N_s=0.270\, ,\quad a_s=0.117\, ,\quad A_s=1.70
\end{eqnarray}
to be compared with the original GRV98 parameters \cite{ref4}:
$N_g=1.443$, $a_g=0.147$, \mbox{$A_g=-2.656\,\,$} and $N_s=0.273$, 
$a_s=0.121$, $A_s=1.80$.  The resulting predictions are compared
to the H1--data \cite{ref2} in Fig.~1.  These results are also 
consistent with the ZEUS--data \cite{ref3} with partly lower
statistics.  The corresponding $\chi^2/dof$ are 0.50 for the 
`best fit' ($dof =48$) and 0.94 for GRV$_{\rm mod}$ ($dof=50$),
respectively.  Our treatment of the heavy flavor contributions
to $F_2$ differs from that in \cite{ref1}.  We evaluate these
contributions in the fixed flavor $f=3$ scheme of \cite{ref4},
together with the massive heavy quark ($c,\, b$) contributions,
rather than in the $f=4$ (massless) scheme utilized in \cite{ref1}.
We have checked, however, that our disagreement with \cite{ref1}
does \underline{not} result from our $f=3$ plus heavy quarks vs.\
the $f=4$ massless quark calculations in \cite{ref1}:  we have 
also performed a fit for $f=4$ massless quarks and the results 
for $F_2$ and its curvature, to be discussed below, remain 
essentially unchanged. 

In Figs.~2 and 3 we show our gluon and sea input distributions in
(2) and (3), as well as their evolved shapes at $Q^2=4.5$ GeV$^2$ in
the small--$x$ region.  It can be seen that both of our new 
small--$x$ gluon distributions at $Q^2=4.5$ GeV$^2$ conform to the
rising shape obtained in most available analyses published so
far, in contrast to the valence--like shape obtained in \cite{ref1}
where the gluon density $xg$ decreases as $x\to 0$.  It is possible to
conceive a valence--like gluon at some very--low $Q^2$ scale, as
in \cite{ref4}, but even in this extreme case the gluon ends up as
non valence--like at $Q^2>1$ GeV$^2$, in particular at $Q^2=4.5$
GeV$^2$, as physically expected.

Turning now to the curvature test of $F_2$ advocated and discussed
in \cite{ref1}, we first present in Fig.~4 our results for $F_2(x,Q^2)$
at $x=10^{-4}$, together with two representative expectations of
global fits \cite{ref5,ref6}, as a function of \cite{ref1}
\begin{equation}
q = \log_{10}\left(1+\frac{Q^2}{0.5\,{\rm GeV}^2}\right)\, .
\end{equation}
This variable has the advantage that most measurements lie along a
straight line \cite{ref1} as indicated by the dotted line at 
$x=10^{-4}$ in Fig.~4.  The MRST01 parametrization \cite{ref5}
results in a sizable curvature for $F_2$ in contrast to all other
fits shown in Fig.~4.  This large curvature, incompatible with 
the data presented in \cite{ref1}, is mainly caused by
the valence--like input gluon distribution of MRST01 at $Q_0^2=1$
GeV$^2$ in the small--$x$ region which becomes even negative for
$x<10^{-3}$ \cite{ref5}.  A similar result was obtained in 
\cite{ref1} based on a particular gluon distribution $xg(x,Q^2)$
which 
decreases with decreasing $x$ for 
$x$ \raisebox{-0.1cm}{$\stackrel{<}{\sim}$} $10^{-3}$ 
even at $Q^2=4.5$ GeV$^2$ (cf.~fig.~7 in \cite{ref1}).
More explicitly the curvature can be 
directly extracted from
\begin{equation}
F_2(x,Q^2)=a_0(x)+a_1(x)q+a_2(x)q^2\, .
\end{equation}
The curvature $a_2(x)=\frac{1}{2}\,\partial^2_q F_2(x,Q^2)$ is
evaluated by fitting the predictions for $F_2(x,Q^2)$ at fixed 
values of $x$ to a (kinematically) given interval of $q$.
In Fig.~5(a) we present $a_2(x)$ which results from experimentally
selected $q$--intervals \cite{ref1}:
\begin{eqnarray}
0.7 \leq q\leq 1.4 & {\rm for} & 2\times 10^{-4}
        < x < 10^{-2}\nonumber\\
0.7 \leq q\leq 1.2 & {\rm for} & 5\times 10^{-5} < x \leq
           2\times 10^{-4}\nonumber\\
0.6 \leq q\leq 0.8 & {\rm for} & x = 5\times 10^{-5}\, .
\end{eqnarray}
Notice that the average value of $q$ decreases with decreasing $x$
due to the kinematically more restricted $Q^2$ range accessible
experimentally. For comparison we also show in Fig.~5(b) the curvature
$a_2(x)$ for an $x$--independent fixed $q$--interval
\begin{eqnarray}
0.6\leq q\leq 1.4\quad\quad
  (1.5\,{\rm GeV}^2\leq Q^2\leq 12\, {\rm GeV}^2)\, .
\end{eqnarray}
Apart from the rather large values of $a_2(x)$ specific for the 
MRST01 fit as discussed above (cf.~fig.~4), our `best fit' and 
GRV$_{\rm mod}$ results, based on the inputs in (4) and (5), 
respectively, do agree well with the experimental curvatures as
calculated and presented in \cite{ref1} using H1 data.  It should be
noted that perturbative NLO evolutions result in a {\underline{positive}}
curvature $a_2(x)$ which increases as $x$ decreases.  This feature
is supported by the data shown in fig.~5(a); since the data point
at $x<10^{-4}$ is statistically insignificant, future precision
measurements in this very small $x$--region should provide a 
sensitive test of the range of validity of perturbative QCD
evolutions.

Furthermore, the H1 collaboration \cite{ref2} has found a good
agreement between the perturbative NLO evolution and the slope of
$F_2(x,Q^2)$, i.e.\ the {\underline{first}} derivative 
$\partial_{Q^2}F_2$. 

To conclude, the perturbative NLO evolution of parton distributions
in the \mbox{small--$x$} region is compatible with recent
high--statistics measurements of the $Q^2$--dependence of $F_2^p(x,Q^2)$
in that region.  A characteristic feature of perturbative QCD
evolutions is a {\underline{positive}} curvature $a_2(x)$ which
increases as $x$ decreases (cf.~fig.~5).  Although present data
are indicative for such a behavior, they are statistically insignificant
for $x<10^{-4}$.  Future precision measurements and the ensuing
improvements of the determination of the curvature in the very
small $x$--region should provide further information concerning
the detailed shapes of the gluon and sea distributions, and
perhaps may even provide a sensitive test of the range of validity
of perturbative QCD.
\vspace{0.5cm}

This work has been supported in part by the `Bundesministerium f\"ur
Bildung und Forschung', Berlin/Bonn. 
\vspace{0.5cm}


\newpage


\begin{figure}
\begin{center}
\epsfig{figure= 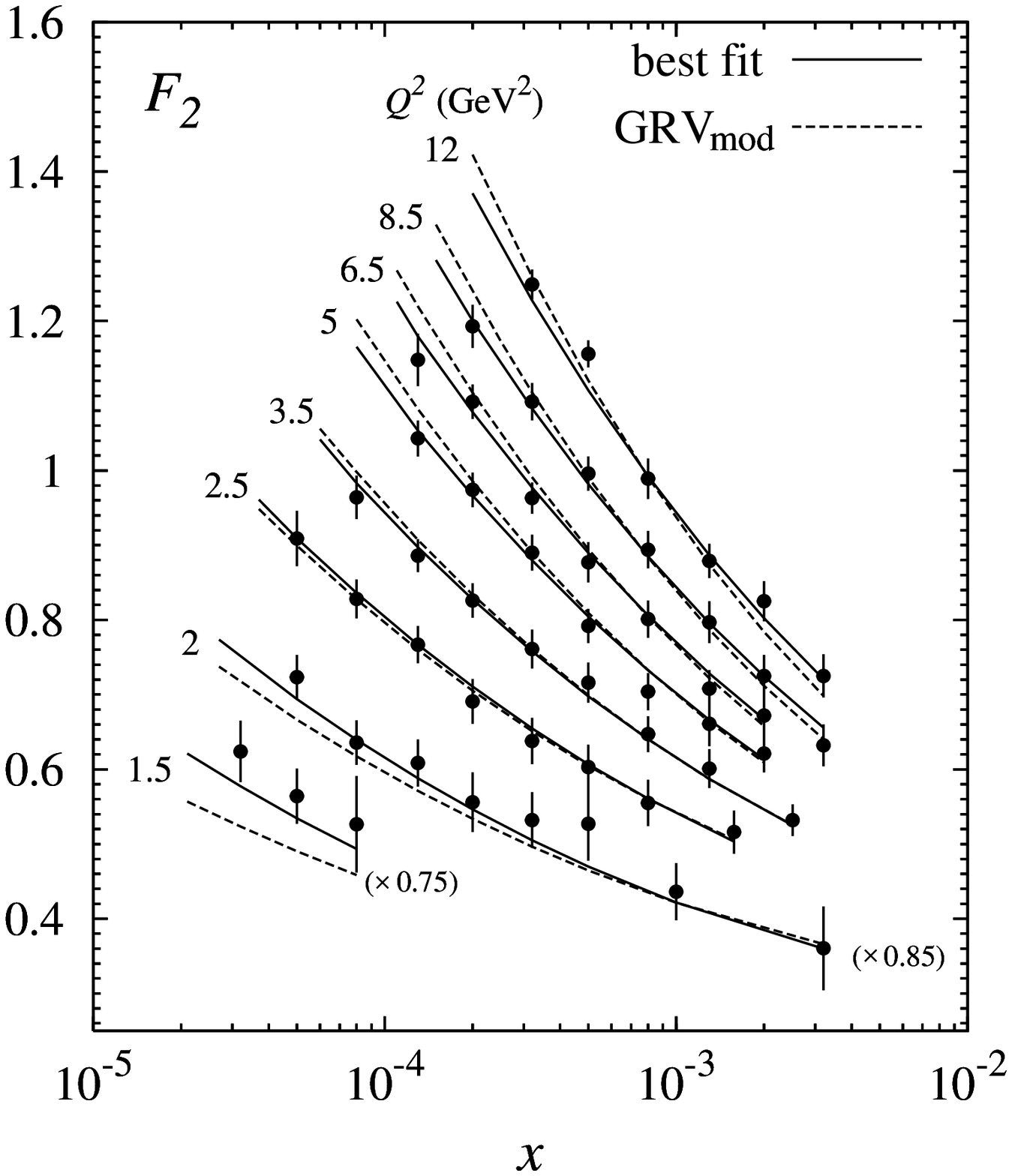, width = 16 cm,height = 15 cm}  
\end{center}
\caption{Comparison of our `best fit' and GRV$_{\rm mod}$
      results based on (4) and (5), respectively, with the H1 data
      \cite{ref2}. To ease the graphical representation, the results
      and data for the lowest bins in $Q^2=1.5$ GeV$^2$ and 2 GeV$^2$
      have been multiplied by 0.75 and 0.85, respectively, as 
      indicated}
\end{figure}
\newpage
\begin{figure}
\begin{center}
\epsfig{figure= 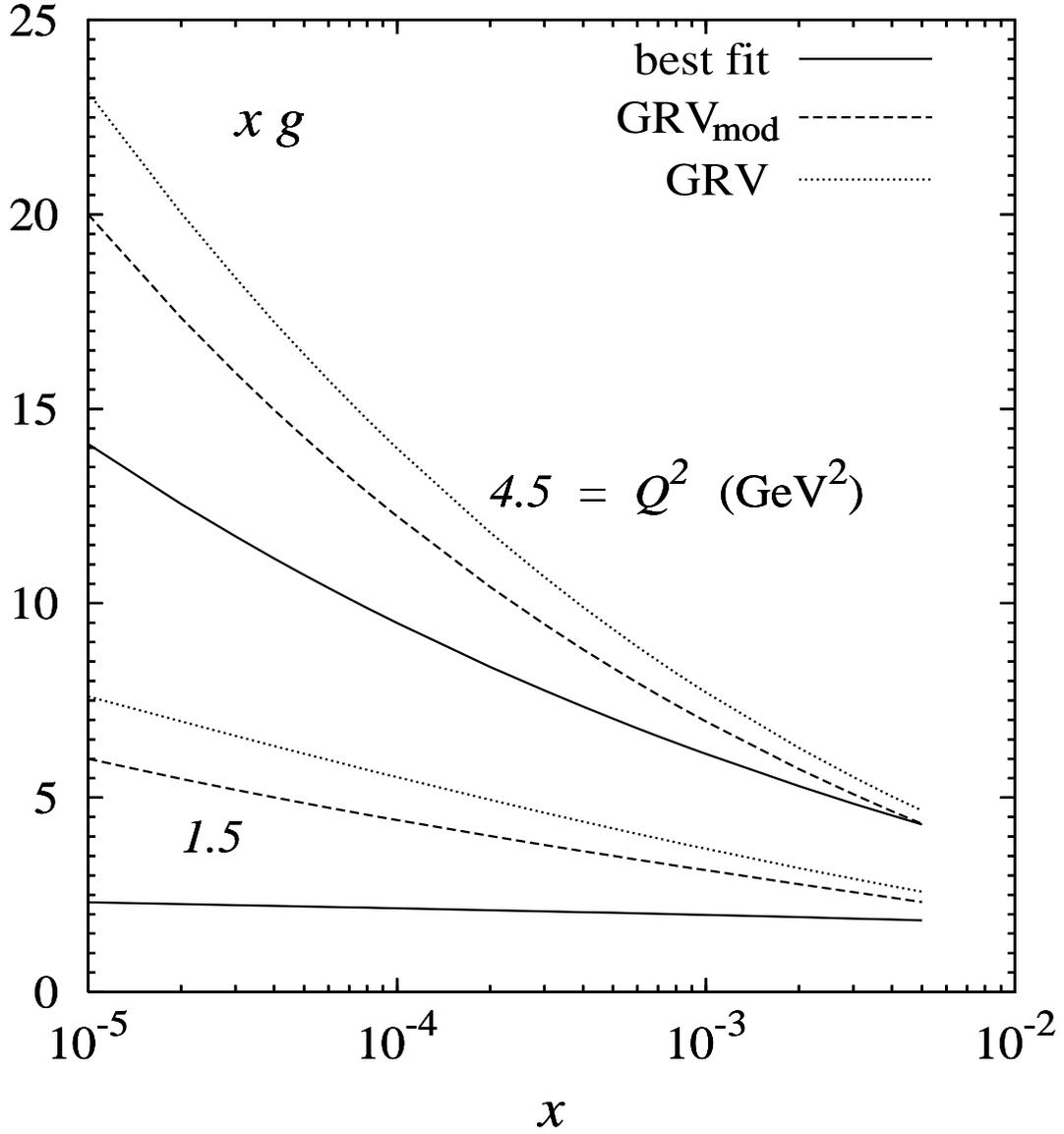, width = 16 cm,height = 15 cm}  
\end{center}
\caption{The gluon distributions at the input scale 
      $Q_0^2= 1.5$ GeV$^2$ corresponding to (2) with the `best fit'
      and GRV$_{\rm mod}$ parameters in (4) and (5), respectively,
      and at $Q^2=4.5$ GeV$^2$.  For comparison, the original GRV98
      results \cite{ref4} are shown as well by the dotted curves}
\end{figure}
\newpage
\begin{figure}
\begin{center}
\epsfig{figure= 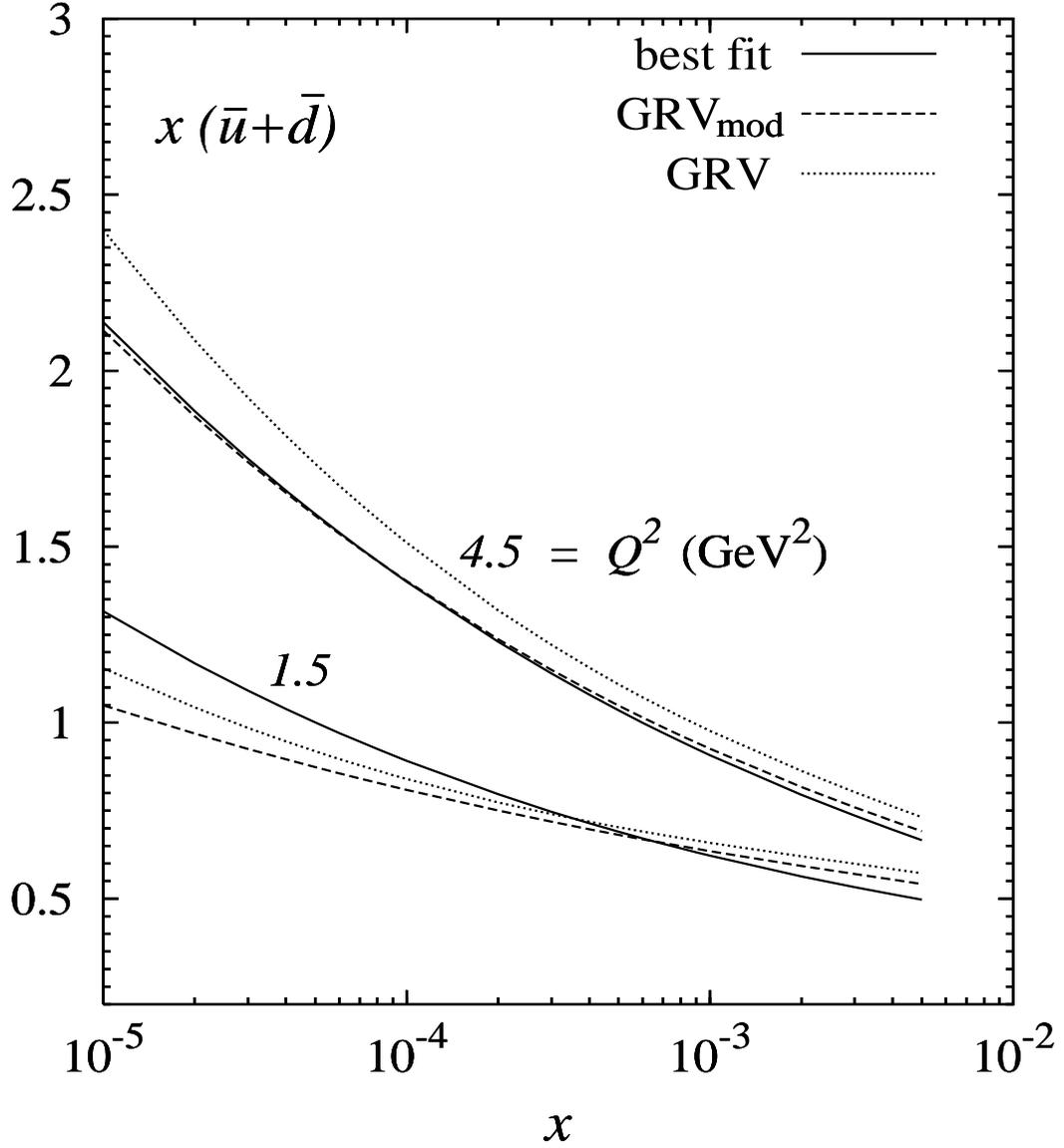, width = 16 cm,height = 15 cm}  
\end{center}
\caption{The sea distribution $x(\bar{u}+\bar{d})$ at
      the input scale $Q_0^2=1.5$ GeV$^2$ in (3) with the `best fit'
      and GRV$_{\rm mod}$ parameters in (4) and (5), respectively,
      and at $Q^2=4.5$ GeV$^2$.  For comparison, the original GRV98
      results \cite{ref4} are shown as well by the dotted curves}
\end{figure}

\newpage
\begin{figure}
\begin{center}
\epsfig{figure= 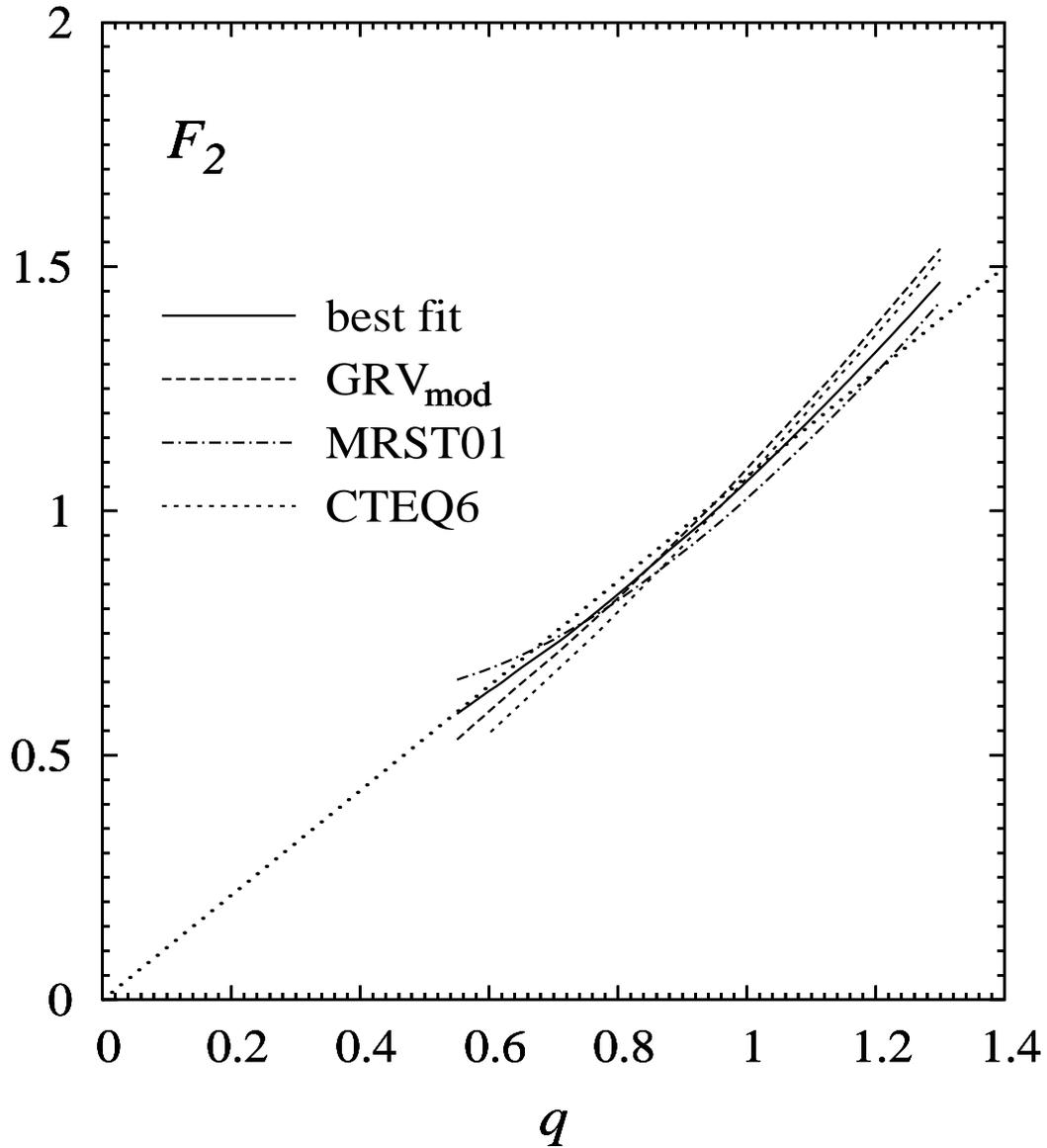, width = 16 cm,height = 15 cm}  
\end{center}
\caption{Predictions for $F_2(x,Q^2)$ at $x=10^{-4}$
      plotted versus $q$ defined in (6).  Representative global fit
      results are taken from MRST01 \cite{ref5} and CTEQ6M \cite{ref6}.
      Most small--$x$ measurements lie along the straight (dotted)
      line \cite{ref1}}
\end{figure}

\newpage
\begin{figure}
\begin{center}
\vspace*{-4cm}
\epsfig{figure= 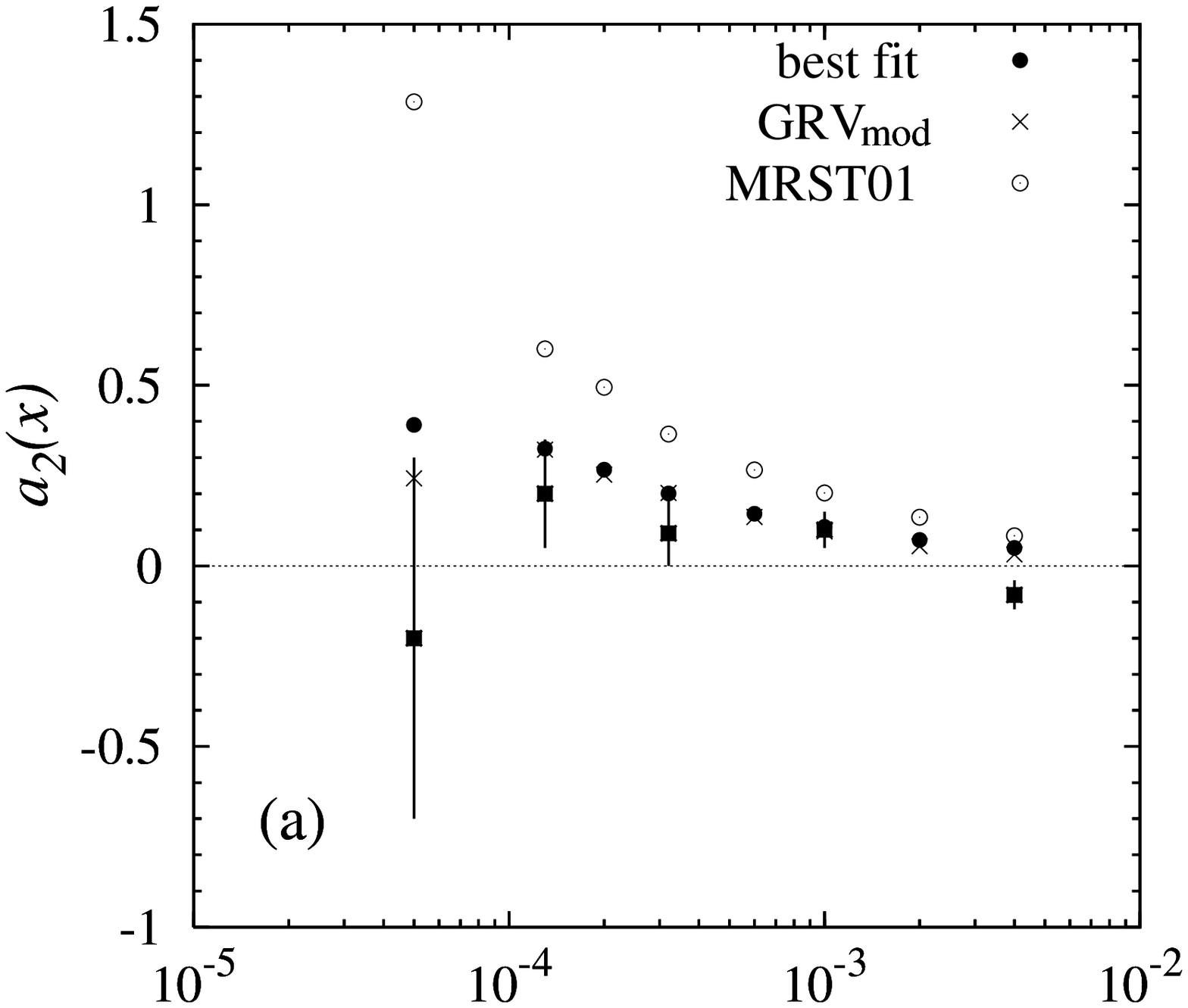, width = 13 cm,height = 10 cm}
\epsfig{figure= 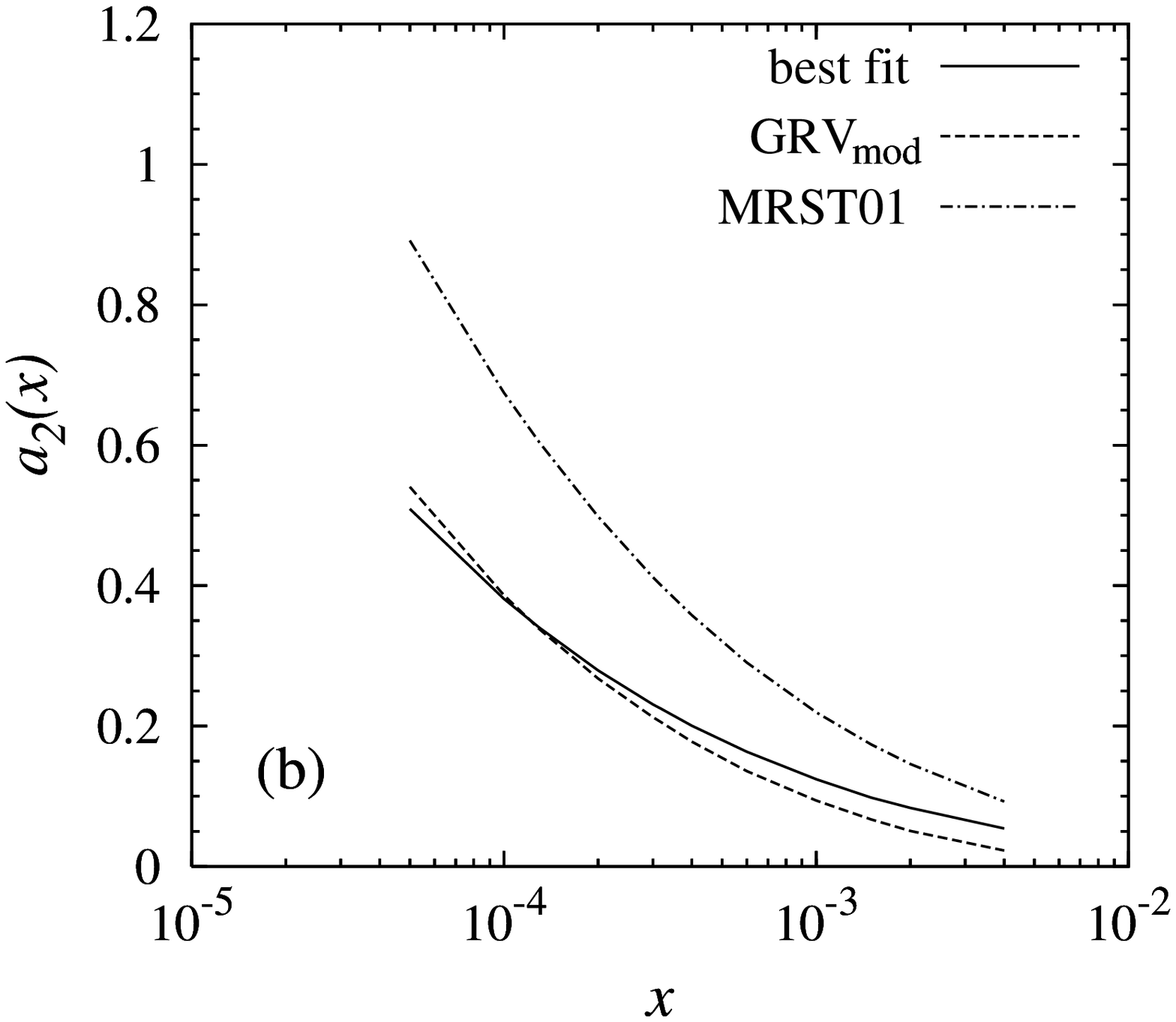, width = 13 cm,height = 10 cm}
\end{center}
\caption{The curvature $a_2(x)$ as defined in (7) for (a)
      the variable $q$--intervals in (8) and (b) the fixed $q$--interval
      in (9).  Also shown are the corresponding MRST01 results
      \cite{ref5}.  The experimental curvatures (squares) shown in (a) are
      taken from \cite{ref1}}
\end{figure}

\end{document}